\documentclass[11pt,a4paper,notitlepage]{article}
\usepackage[left=2.5cm, right=2.5cm, top=2.5cm, bottom=2.5cm]{geometry}
\usepackage{amssymb}
\usepackage{float}
\usepackage{graphicx}
\usepackage{amsmath}
\usepackage{booktabs} 
\usepackage{makecell} 
\usepackage[backend=biber, style=nature, doi=true, url=false]{biblatex}
\title{\bf {Risk Theory and Pricing of  "Pay-for-Performance" Business Models}}
\author{Roger Knecktys$^1$, Henrik Bette$^1$, Rüdiger Kiesel$^2$, Thomas Guhr$^1$\\
$^1$\small{\textit{Fakultät für Physik, Universität Duisburg-Essen, Duisburg, Germany}}\\
$^2$\small{\textit{Fakultät für Wirtschaftswissenschaften, Universität Duisburg-Essen, Essen, Germany
}}}
\date{}
\begin{document}
\maketitle
\begin{abstract}
Technology trends as digitalization and Industry 4.0 initiate a growing demand for new business models. Most of this models requires a fundamental shift of operational and financial risks between seller and buyer. A key question is therefore how  to include additional risk pricing and hedging. In this paper we propose a new approach for a risk theory of innovative performance based business models as "Pay-for-Performance" or "Product as a Service". A new model and calculation method for determination the risk premium is presented. It contains beside financial price fluctuations also operational failure behavior of products. We apply the model for a typical industrial application and simulate the pricing dependency for different cost distributions.   
\end{abstract}
\newpage
\section {Introduction} 
New and innovative business models play an eminent role for future business in addition to the known technology trends as digitalization. Technology alone cannot answer the key questions for the end user: “What is the basic value proposition and how can I use it to improve my business.” Hence a business model is needed to transfer technology into real competitive advantages. In many branches the end user is not really interested in buying and operating a product but only interested in the outcome or the performance of the underlying technology. Theodore Levitt's known cite is: “People don´t want to buy a quarter-inch drill. They want a quarter-inch hole!” \cite {Deloitte2021}. In the case of medical imaging systems the end user is not interested in the complex technology of a scanner rather the ability to produce pictures of predefined quality is relevant.
\begin{figure}[H]
\centering
	\includegraphics[width=0.90\textwidth]{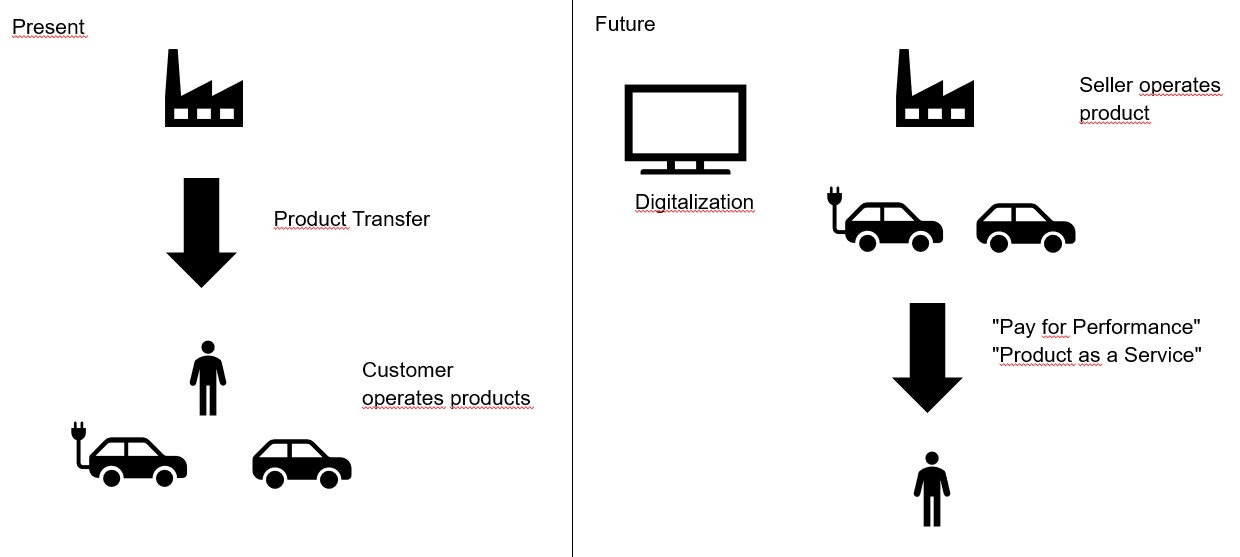}
	\caption{Future of Business}
\end{figure}
From this perspective a technology partnership which can provide service and operation of a complex system seems worthwhile. The end user will only pay a flat payment depending on the performance or output – this is called “Pay-For-Performance” or "As a service business" \cite{Schnaars2022}. In economic terms a typical product business is transformed to a pure service business model. A fundamental risk transfer takes place from the end user to the service provider. 

The advantage is obvious. In healthcare the (medical) competence of the end user is often very disparate from the product technology. Thus  managing the installed base or apparatus is an effort larger for the user than for the system provider. Also for the service and technology provider this business models have certain advantages. Due to global competition the accessible profit just for commodity products is strictly limited. The new business models combine product business with product related services and moreover with IT service and consulting. With this mix of service and product business a significant increase of profit is possible but of course with complete new types of risks involved. This cross product-service business increases the margin and is especially interesting for large global acting companies with high overhead cost. A second advantage is that technology partnerships are typically long-term contracts (up to 10 years). In this period both partners are shielded from third party competition which gives planning reliability. The management of such contracts has to deal with two different sources of risks which need to be managed and hedged: Financial risk which is caused by price fluctuation of the goods. Operational risk which is caused by the technical availability of the production infrastructure. As common in financial mathematics we can model such price fluctuations with stochastic processes. Short term continuous fluctuations can be modeled with a Brownian Motion as discontinuous jumps with a Compound Poisson process. The general class of this combined processes are the well-known Lévy processes. An important task is to develop strategies for hedging both risks in order to run the business in an anticipatory way. For the financial side the development of various derivatives provides the management with appropriate instruments to hedge such risks. Modern financial mathematics lays the ground for the price calculation of such derivatives. Since the paramount results of Black, Merton and Scholes we are in possession of a basic theory how to calculate a “fair” price for these instruments. A key question is whether it is possible to include operational risks into the existing (financial) models or even more fundamental if it makes sense at all to combine these two areas of management. At a first sight these two aspects (financial/technical) has nothing in common. The different risks can be separately managed e.g. options for financial risks, e.g. redundancy for technical risks but for both risks the management needs to calculate a “fair” price. Moreover a combined perspective seems to be unneeded. The ongoing digitalization of the products and their usage known as Industry 4.0 brings now other perspectives into this subject. The classical way of product business (complete transfer of the product ownership to a buyer) will be complemented with new business models as e.g. Performance Based Contracting (PBC) or “Pay-for-Performance”. Here the risk to operate and service the product on behalf of the customer stays with the seller. It is a combination of product and service business. From the buyer point of view it is an outsourcing of the operational risk to the seller. For this transfer of risk the  seller demands compensation for the life cycle cost plus a premium for the risk take-over. The advantage for both sides is obvious. In many branches the operational management of complex products requires technical competencies which the (product) buyer does not have but which is core competence of the product seller. A performance contract gives planning reliability for the buyer and unburdened from (operational) risk management tasks. Resources can be better allocated used for their core business. In this paper we present a comprehensive model for PBC and simulate the risk premium for a simplified example based on real data.  
\begin{figure}
	\includegraphics[width=0.90\textwidth]{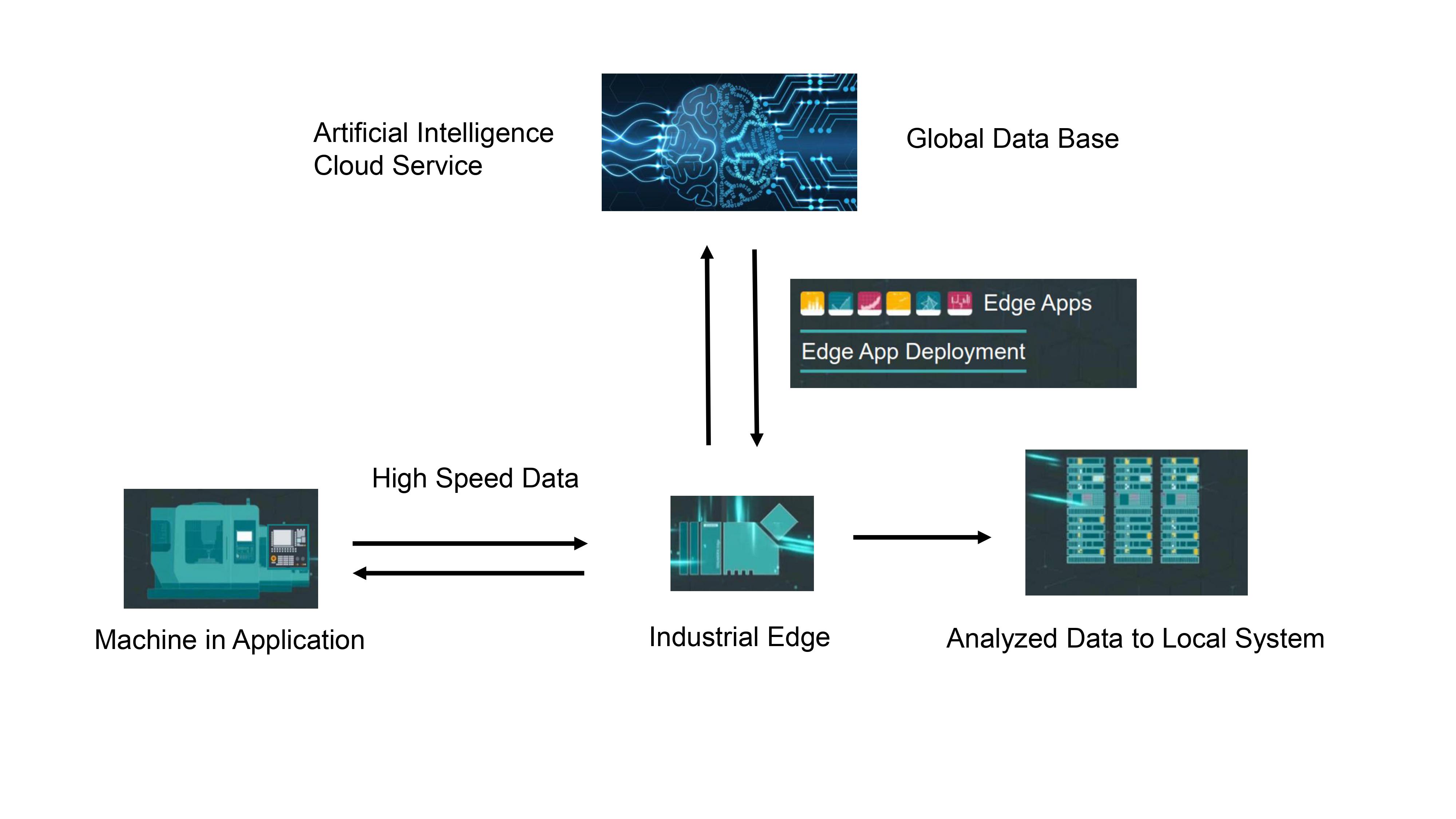}
	\caption{Product Connectivity and Data Aquisition}
	\label{fig:onlinedata}
\end{figure}

The paper is organized as follows: In Sec.2 of the paper we define the notion "Performance Based Contract" in the framework of fundamental asset pricing and insurance theory. In Sec.3 the general model for PBC is defined. It is based on a Lévy process which includes the cost jumps of the life cycle cost beside of the Brownian diffusion. In Sec.4  we extend the no-arbitrage argument of asset pricing to a fair pricing in risk-sharing models. Then we apply the standard price calculation as the expectation value of the pay-off functions. In Sec. 5 we present Monte Carlo simulations for different probability jump distributions. We use then empirical cost in order to determine realistic cost distributions. Based on the data we simulate the risk premium and pricing.
\section{Definition and Examples of Performance Contracts}  
The notion of "Performance Contracting" is not well defined.   Most of these new business models are not standardized. Also the term "Business as a service = XaaS" is more a general concept for contracting and has to be specified with economical terms and conditions. 
The following example should show what are common key elements of such contracts and how they are linked to inherent risks.      
\newline
\\
\textbf{Definition:} \textit{A price model for a Performance Based Contract (PBC) is a function $P(X_t,h(X_t,K))$ over a stochastic cost variable $X_t$ and a payoff function $h(X_t,K)$ which is dependent on the values of predefined "Key Performance Indicators".}
\newline
\\
Typical examples for "Key Performance Indicators" can be cost values, up-time/downtime, energy cost etc. The definition is comparable to other financial derivatives e.g. barrier options in which the pay-off is dependent on the path of the stochastic variables and defined parameters \cite{Korn2010}\cite{Musiela2005}.
 Examples of this key performance indicators are service cost, energy consumption, operating hours or delivered units. In a performance contract one may agree that payments are only due if the value of $X_t$ is exceeding or fall below a certain limit during the contract period.  
 
 A typical example of a PBC is a manufacturer of turbines for airplanes which are sold to a flight company. The turbine company offers traditionally the product which is integrated into the system and additional service and maintenance. As the turbines are key elements of the overall system the supplier of turbines play an eminent role for the reliability of the flight company.Hence the idea is to negotiate a "Performance Contract". The flight company pays a flat fee to the turbine producer only for covered miles including service, maintenance and spare parts (life cycle cost). A part of the operational risk is now transferred to the supplier because the performance measure "covered mileage" is not only dependent on the availability of the turbines but also on the live load. In all performance contracts a "Key Performance Indicator"KPI has to be defined which is the "basic stochastic underlying" to calculate the contract price. As the KPI is influenced by different types of external variables data of time series are essential in order to manage the financial risk. It is immediate clear that this business models are directly linked to a shared data availability and analysis. At the same time an alignment of mutual business goals over the supply chain is reached. The turbine supplier as well as the flight company are interested in high utilization of the components. 
So also the delivery of a good product quality is enforced.  
In this example the payment is linked to the risk of external market development which is not easy to manage. Another favorite example of a KPI  for such contracts is energy consumption. The provider guarantees a certain reduction in energy cost and use the money to reimburse the product investment. Here the KPI is linked to energy pricing. 
In our application the contract is limited to the coverage of fluctuating life cycle cost. These cost can be easily measured and modeled by established financial models as Compound Poisson Process, Wiener Process or in general with Lévy Processes and are focus of this paper.\cite{Merton1973} The data we are using are life cycle cost for electrical drive systems which contains electrical motors and frequency converters. This kind of drives are present in many different industrial branches as oil/gas, water, transport system etc.   
\section {Fundamental Model of Performance Contracts}
For a PBC the buyer and seller enter a contract which guarantees a certain defined performance measured on real data. Typical for this agreements is a flat payment of the buyer over a defined time that covers all service and operating cost for the installed base. In general, we assume that the life cycle cost can be modeled by a general Lévy Process which is a combination of continuous cost fluctuation (e.g. energy cost fluctuation) and service cost as jumps.   
\begin{equation}
X_t=\gamma t + \sigma W_t+\sum_{i=0}^{N(t)} J_{i}  
\end{equation} 
The first term $\gamma$ gives the drift which is a linear price trend, the $W_t$ is a Wiener Process for small stochastic fluctuations and the last term is a compound Poisson process for model are the dominant jumps of the service cost.
In contrast to classical assumption we are not using geometric stochastic process as our empirical data are absolute cost values \cite{Bingham2004} \cite{Ziemann2021} \cite{Hull2006} \cite{Ruttiens2013}\cite{Ingber2000}.The Figure 3 shows a typical path of a Lévy process which describe larger jumps (service cost) and smaller stochastic price fluctuations (fluctuating consumption cost)  
\begin{figure}[H]
	\includegraphics[width=0.90\textwidth]{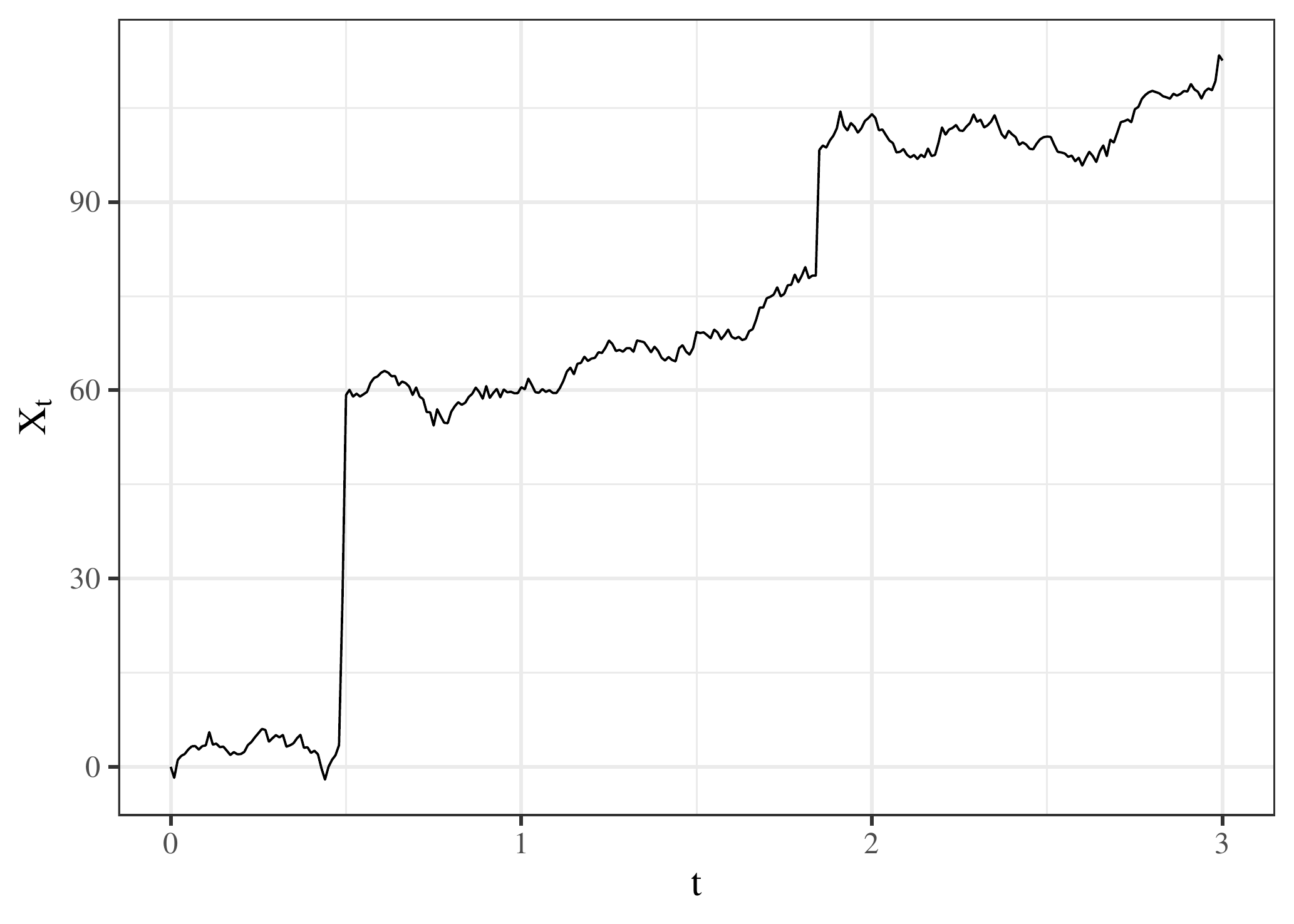}
	\caption{Example of a Lévy process sample path}
	\label{fig:samplepathlevy}
\end{figure}
We use an inhomogeneous Compound Poisson Process to describe a failure counting process with different jump heights. We set the jump rate of our cost function with an empirical distribution of the jump heights given by empirical data.   
\begin{equation}
\mathbb {P}(N_t-N_{s}=k)=e^{-\lambda(t-s)}\frac{(\lambda(t-s))^k}{k!}, \;\; k\epsilon\mathbb{N} 
\end{equation}
with $0\leq s \leq t$, $\lambda$ is the failure rate of the system and we assume in the general case that
$\lambda$ can be a deterministic function of time $\lambda$(t) then
\begin{equation}
\Lambda(t)=\int\limits_{0}^{t}\lambda(t')dt'
\end{equation}
is the cumulated failure rate. $\lambda$(t) and $J(t)$ have to be determined from empirical real data for the underlying technical products 
\begin{figure}
	\includegraphics[width=0.90\textwidth]{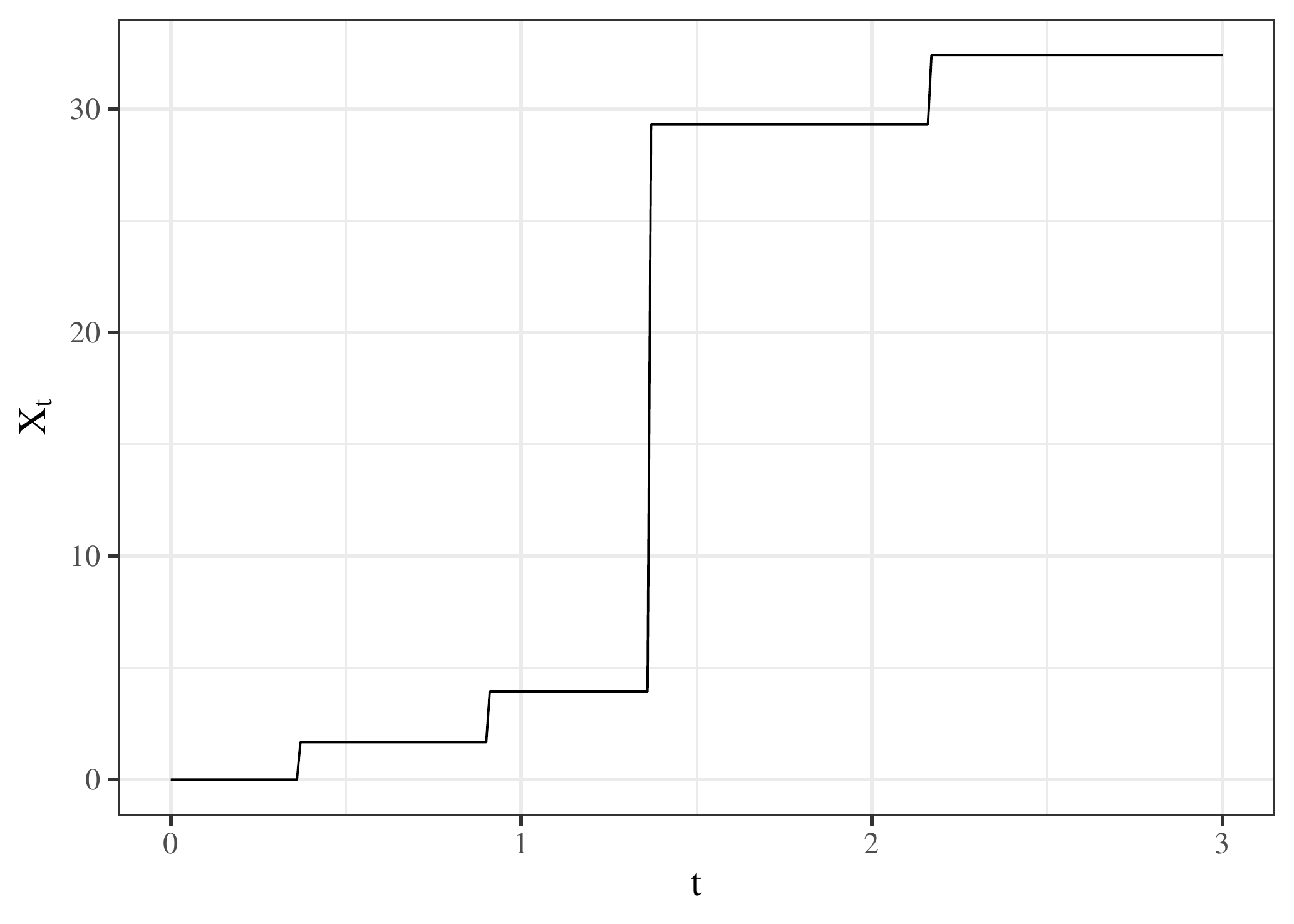}
	\caption{Example of a compound Poisson Process}
	\label{fig:samplepath}
\end{figure}
The figure shows a simple path of a jump process with cost jumps which are statistically independent and evenly distributed. \cite{Kwok2008}    
\section {Risk Theory and Contract Pricing}
In order to continue with a pricing model we need some pre-consideration concerning risk theory of such contracts. We start with a very general definition that is commonly used for the definition of a financial risk. \cite{Schmidli2017}
\newline
\\
\textbf{Definition: }\textit{A financial risk is the deviation from an expected value of an underlying stochastic price process.}
\newline
\\
The taking over of a financial risk between market participants should be compensated by a risk premium. There are different possibilities to define a price for such a risk premium.The natural choice to calculate the risk premium is the expectation value of the discounted payoff for a given probability distribution of the underlying price process.\cite{Voit2005} 
\begin {equation} 
C_{\tau}=e^{-\gamma\tau}\mathbb{E}_{Q}[h(X_T)]
\end{equation}
with the expectation value operator $\mathbb{E}_{Q}$ defined over a probability space $(X,Q,\Omega)$ as
\begin {equation} 
\mathbb{E}_{Q}=\int\limits_\Omega X dQ
\end{equation}
This price definition is very general and needs to be specified for given market conditions (tradable good, efficient market, risk transfer). The customer-realized- price is usually different to this expectation value. In the case of the Black-Scholes model which assumes an arbitrage free market with efficient tradable goods the price for an option can be calculated by 
\begin {equation} 
C_{\tau}=e^{-r\tau}\mathbb{E}_{Q*}[h(X_T)]
\end{equation}
The trend parameter $\gamma$ is replaced by the risk free interest rate r. The probability distribution is changed from Q to Q* which is a martingale measure reflecting the no-arbitrage condition. This result can also be obtained from portfolio
models and using the Feynman-Kac formula \cite {Bingham1998}\cite{shreve2004stochastic}. In the case of a PBC the underlying asset is service cost, which is not tradable. But we will see in chapter 4.2  that we can use the risk-neutral pricing method in the context of a fair risk-sharing between the contract partners.
\subsection {Price of Insurance Contracts}        
For an insurance contract with stochastic claims $X_i$  the price for the premium can be defined in different ways in order to ensure that the initial capital of the insurer will not be negative (ruin problem) so the most easiest choice for pricing will be just the expectation value of $X_i$ with a certain safety loading $\theta$ 
\begin{equation}
F=(1+\theta)\mathbb{E}[X_t]
\end{equation}
The parameter $\theta$ represents the role of an expected profit and is related to the drift parameter $\gamma$ in the option pricing methodology.
The disadvantage of this definition is clear: It is not sensitive to fat tail distribution of the claims X. In our case the stochastic claims $X_t$ are mainly driven by the jumps of the life cycle cost in the contract. Thus a natural extension for the price model is adding an additional term which depends on the variation 
of the claim process \cite{Schmidli2017}
\begin{equation}
	F=\mathbb{E}[X_t]+\alpha \text{Var}[X_t]
\end{equation}
with a parameter $\alpha>0$ which is arbitrary. In our model we use a fair option pricing for the variance depending risk premium. Insurance risk theory is more focused on ruin probability than on pricing issues. As the second term depends on the volatility of the life cycle cost it includes an additional risk premium for heavy tailed cost distributions. As we will see in chapter 3.2 real life cycle cost data shows are highly asymmetric and heavy tailed. The principle flaw of actuarial calculations as in equation [8] is that the dependency from the variance of the claims $\alpha$ is arbitrary. The question here is whether we can use basic ideas from risk-neutral valuation for option pricing in order to define a fair price. As we have no tradable goods in this kind of contracts we need a different definition of fair pricing. In our case it is not an equilibrium state which is generate by an efficient market. It is the result of the negotiation process of the contract partners.  
\begin{table}
	\centering
	\caption{Risk profiles of financial instruments}
	\label{fig:hedginginssurance}
	\begin{tabular}{llll}
		\toprule
		Type of contract & Risk Management & Probabilities & Methods \\
		\midrule
		\midrule
		Insurance & Shift risk & \makecell[l]{Low probability \\ high financial cost} & \makecell[l]{Extreme value \\ theory} \\
		\midrule
		Hedging & Off-set risk & \makecell[l]{Medium/High probability \\ medium cost} & \makecell[l]{Geometric Brownian \\ Motion} \\
		\midrule
		Performance contract & Shift risk & \makecell[l]{Medium/High probability \\ medium cost} & \makecell[l]{Extreme fat tail \\ distribution} \\
		\bottomrule
	\end{tabular}
\end{table}        
\subsection{Price for Performance Contracts}
Our idea is to combine the calculation of the risk premium with the idea of risk neutral valuation from option pricing in order to define a fair price which then depends on the volatility.
In principle an option price $C$ based on a underlying price process $X_t$ can be defined as an expectation value over the payoff function $h(X,K)$ with the probability distribution $\psi(X,t)$ of $X_t$ and the strike price $K$. The strike price is the predefined price at the due date of the contract. 
\begin{equation}
C=e^{-\gamma t}\int\limits_K^\infty h(X,K)\Psi(X,t)dX
\end{equation} 
The formula has a very intuitively explanation as the option price is the expectation value of the pay-off function and "discounted" with the drift term. 
 As already mentioned in \cite{Bingham2004} we can write in general
\begin{equation}
C_{\tau}=e^{-\gamma\tau}\mathbb{E}_{\Psi}[h(X_T)] 
\end{equation}
with $\tau=T-t$ . The probability distribution $\Psi$ can be derived in principle from a Fokker-Planck equation which gives the time evolution of the probability density function $\Psi$ of the stochastic process. 
Only in special cases (e.g. Black-Scholes) this can be solved analytically. In practice Monte-Carlo-Simulations are used to gain numerical results. For the risk neutral valuation the probability measure is changed to a new measure $\Psi^{*}$  hereby the drift parameter  $\gamma$ is changed to the risk-free return rate r.
which is called an "Equivalent Martingale Measure".  
This leads to a change of the probability distribution to an equivalent measure so that the process will be a martingale which reflects the postulate for a  fair price. 
With this argument In the case for efficient markets this will be realized to an assumed equilibrium state because the trading will abolish any arbitrage possibility. In the case PBC the argument is different as we have not tradable assets and financial derivatives. Nevertheless we adapt the no-arbitrage principle in the sense that we share the risk between the two parties in such a way that we gain a fair distribution. 
The central definition reads
\\
\newline
\textbf{Definition of Fair Risk Sharing:} \textit{A risk sharing agreement is called fair  if the underlying process $X_t$ is a martingale.}
\\         
\newline
The definition simply says that both parties should have the same probability to make profit or loss.
The fundamental theorem of asset pricing combines the postulate of no-arbitrage in an efficient market with the existence of an equivalent martingale measure. Thus the market dynamic with an assumed equilibrium state forces the stochastic process to be a martingale. In our case this dynamic is substituted by an agreement between the two contract parties on fair risk sharing. The equilibrium is not the result of a market behavior with many agents. It is an assumption about rational choice.   
In general, the real economic pricing process is never pure rational. The realized price in a market is of course always different to models. In comparison to physics also many assumptions in economics are presumed and not experimentally verifiable. But a good model should at least give a first reasonable approach for describing economical processes.        
\\
In order to apply the above principle for a practical example we neglect in the Lévy process the drift term $\gamma=0$ and the Wiener part $W_t=0$. The extension with non-vanishing drift term is straight forward. But in contrast to the classical Black-Scholes Theory the interpretation of the discount factor will be different. As we can use the actual price of the product as the numéraire the drift paramater reflects the depreciation of the product price over the time.The Wiener process stands for cost fluctuations assumed to be small in comparison to the jump cost.. Therefore we concentrate on a pure jump process as the simplest Lévy process. This leads to:
\begin{equation}
	X_t =\sum_{i=0}^{N(t)} J_{i}  
\end{equation}
 In contrast to the standard financial assumption we are not using geometric Lévy processes as the absolute jump values are directly given as empirical data.  
According to the definition we define an adapted process $X^{*}_t$ which is a martingale
\begin{equation}
		X^{*}_t =\sum_{i=0}^{N(t)}( J_{i} - \lambda t \mathbb{E}[J_i])
\end{equation}
$\lambda$ gives the failure rate of the process and  $\mathbb{E}[J_i]$ is the expectation value of a single cost jumps.
Thus we can calculate the risk premium with the pay-off function $h(X,K) = \text{max}[0,X-K]$ and we choose as the strike price $K= \mathbb{E}[J_i]$ according [8]
\begin{equation}
	C= \mathbb{E}_{\Psi^{*}}[h(X_t,K)]= \int\limits_{K}^{\infty} h(X_t,K)\Psi^{*}(X_t,t)dX
\end{equation}
The Price $P$ for the PCB is then
\begin{equation}
	P=C+\mathbb{E}_{\Psi^{*}}[X_t]
\end{equation}
The calculation of the adapted probability density function $\Psi^{*}$ is not possible in closed form. It can be solved numerically with Monte-Carlo simulations.
\section{Monte Carlo Simulations and application with empirical data}
First as a simple example we simulate the price for a PBC with a underlying cost process with Gaussian distributed jumps \cite{Korn2010} then we simulate the price with real data from the an industry application for drive systems. It is interesting to see how the price depends on the jump rate $\lambda$. In practice this parameter has to be determined from field data using standard reliability analysis. In general it is time-dependent reflecting the aging of a system. 
For this proof of concept we restrict our simulation to the pure compound Poisson process with constant jump rate and vanishing trend parameter. The price for such contracts can then be derived by  
\begin{equation}
	P=\int\limits_K^{\infty} h(X_t,K)\Psi^{*}(X_t)dX+\mathbb{E}[X_t]
\end{equation} 
The expectation value of a Compound Poisson process is given by
\begin{equation}
\mathbb{E}[X_t]= \lambda T \mathbb{E}[J_i]
\end{equation}
The strike price $K$ is also chosen to be the expectation value of the compound Poisson process. Both parameters: the average jump cost and the average number of jumps (=failure rate) can be derived from empirical data of the products in the field
\begin{equation}
K = \lambda T \mathbb{E}[J_i]
\end{equation}  
\begin{figure}
	\includegraphics[width=0.90\textwidth]{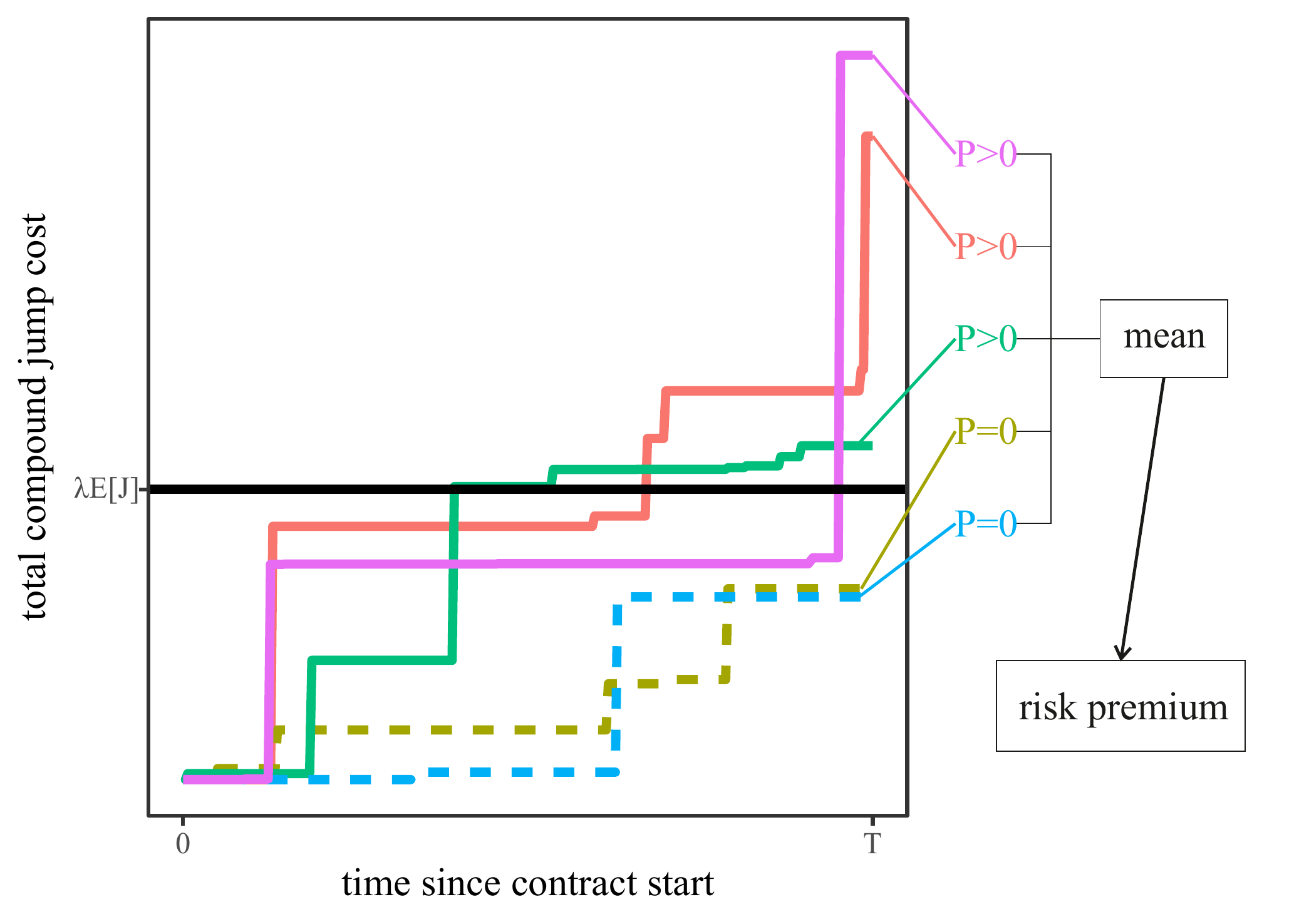}
	\caption{Monte Carlo Path Simulation}
	\label{fig:payoff_skizze}
\end{figure}
The numerical calculation of the premium is based on multiple simulations of the underlying process. As this is a pure jump process the first step is determining the times of jumps based jump rate $\lambda$ per time unit, which follow a Poisson process. Assuming $t_0 = 0$ as start time and $T$ as end time we obtain the times where jumps occur via
\begin{equation}
	t_{i+1} = t_i - \ln(1-\epsilon)/\lambda ~, ~t_i < T,
\end{equation}
where $\epsilon$ is an equally distributed random variable between 0 and 1. The value $X(t)$ of the jump process is then given by
\begin{equation}
	X_t = \sum_{i>0} \Theta(t-t_i) J_i
\end{equation}
with the step function $\Theta$ and the height $J_i$ of jump $i$. In accordance with our service data, the jump heights are simulated as a random variable according to the probability distribution of the jump cost. 
As described we get the final price as the expectation value of the pay-off function over the compensated jump process plus the strike price with the average jump height $\mathbb{E}[J]$. This is calculated analogous to an option price on the financial market from our Monte Carlo simulation. For each simulation run we calculate the payoff with strike price equal to the expectation value $K=\mathbb{E}[X_t]$ and then take the average over all $N$ simulation runs to get the option price $C$.
\begin{equation}
	C = \mathbb{E}[\max(X_T-K, 0)]
\end{equation}
The total price of the PBC is then 
\begin{equation}
	P = C+ \mathbb{E}[X_t]
\end{equation}
Figure 6 shows the additional risk premium $C$ for Gaussian distributed jump cost for different standard deviations.
\begin{figure}
	\includegraphics[width=0.90\textwidth]{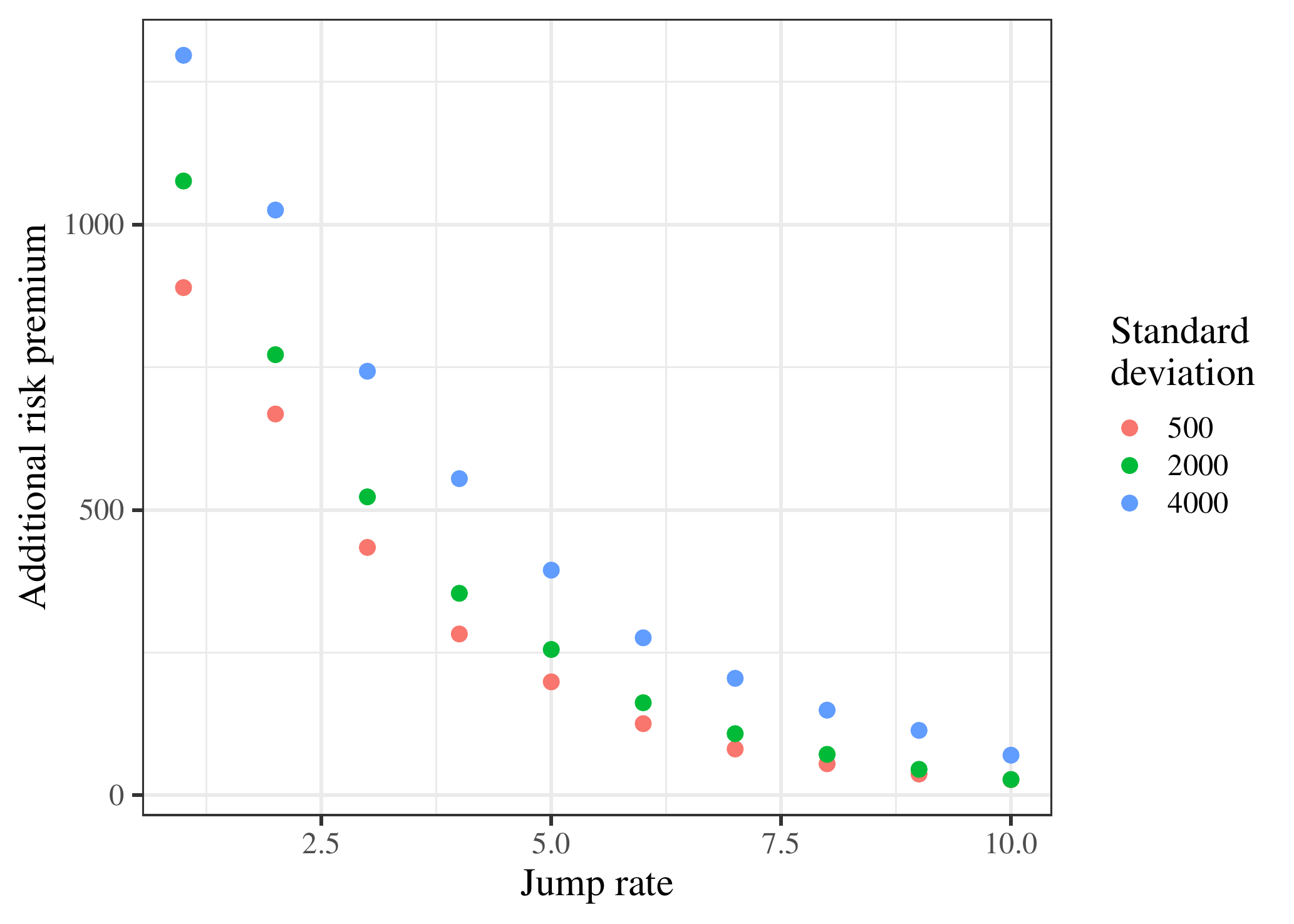}
	\caption{Risk Premium for normal distributed Jump Cost}
	\label{fig:optionPrice_overLambda_norm}
\end{figure}
As expected the risk premium is dependent on the standard deviation of the Gaussian distribution as the measure of the risk. But the value is converging fast to zero with increasing failure rate. It means that the expected cost are a reasonable and fair price if the failure rate is high and the cost are Gaussian distributed. From the graph we can assume that the Risk Premium $C(\lambda]$ is function exp(-$\lambda$). In fact it can be shown \cite {Di_Crescenzo_2015} that the probability distribution $h_J$ of a general compound Poisson process with jump distribution is given by
\begin{equation}
	h_{J}=\sum_{m=1}^{\infty}e^{-\lambda t}\frac{(\lambda t)^{m}}{m!}f^{(m)}_{J}
\end{equation}
Here $f^{(m)}_{J}$ is the m-fold convolution of the jump distribution. As all convolutions from Gaussian functions are again Gaussian the leading term for $m=1$ shows the the exponential behavior. In reality the life cycle cost will be not Gaussian distributed so we simulate an example based on real data for a drive system.       
\subsection{Simulation with empirical data}
The parameters for the distribution of the jump height are taken from a empirical data base for an electrical drive system which consists of motors and frequency converters as main components [Siemens RQPC Data].
\begin{figure}
	\includegraphics[width=0.80\textwidth]{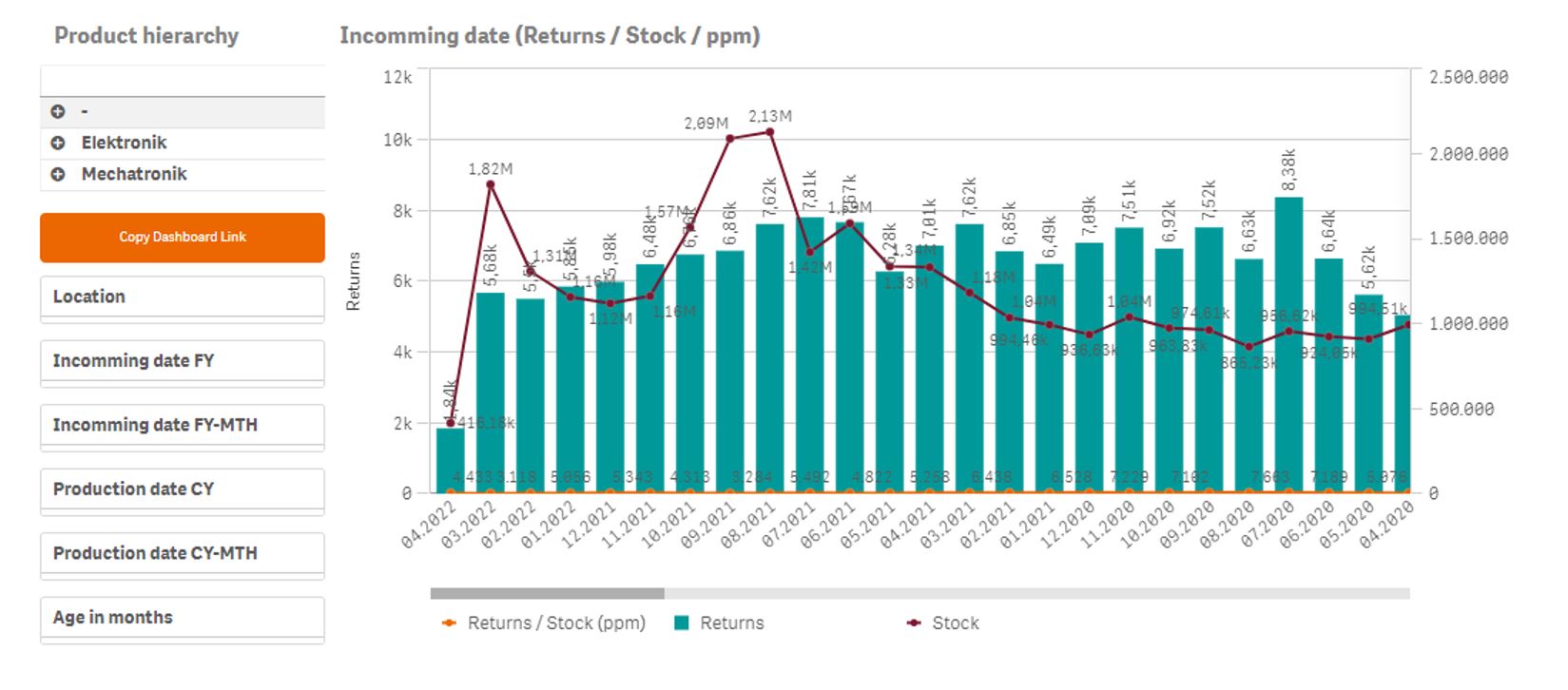}
	\caption{Screenshot Service Database for Life Cycle Information of Drives [10]  }
	\label{fig:RQPCDatabase1}
\end{figure}
This allows the calculation of the premium price for a practical example The service data include all repair cost of the components in different applications for 2 years. As the data are captured within the warranty period of 18 months after production it reflects early failure.
\begin{figure}[H]
	\includegraphics[width=0.80\textwidth]{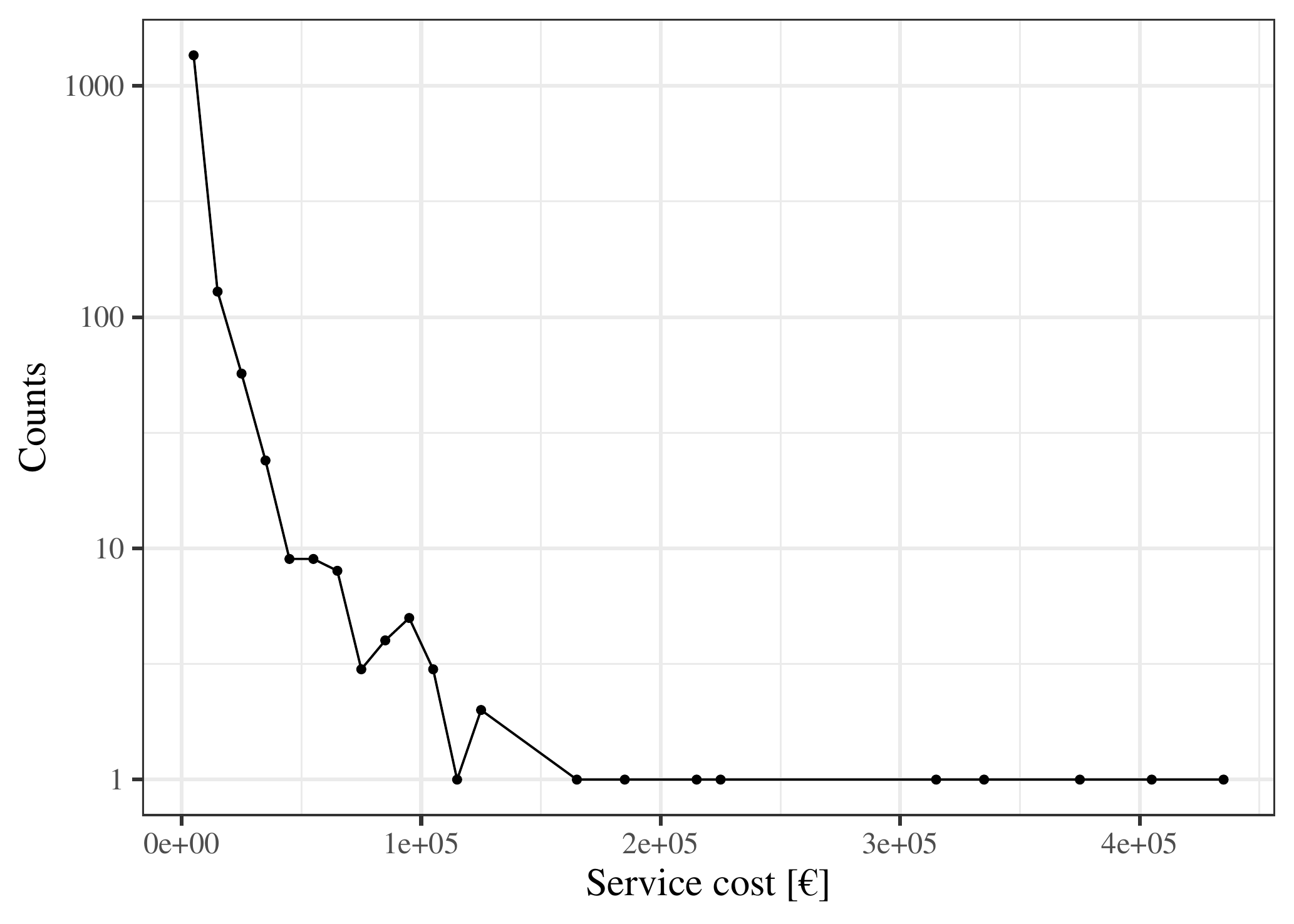}
	\caption{Screenshot Service Database for Life Cycle Information of Drives [10]  }
	\label{fig:cost_hist_pints_filtered_log}
\end{figure}
As Figure 8 shows the spread of the cost distribution is high. Thus we fit the logarithm of the cost with a Weibull distribution   
Figure 9 gives a best fit for a cumulated Weibull Distribution with the indicated parameters. It shows that the variance of the cost is high as expected. 
\begin{figure}[H]
	\includegraphics[width=0.80\textwidth]{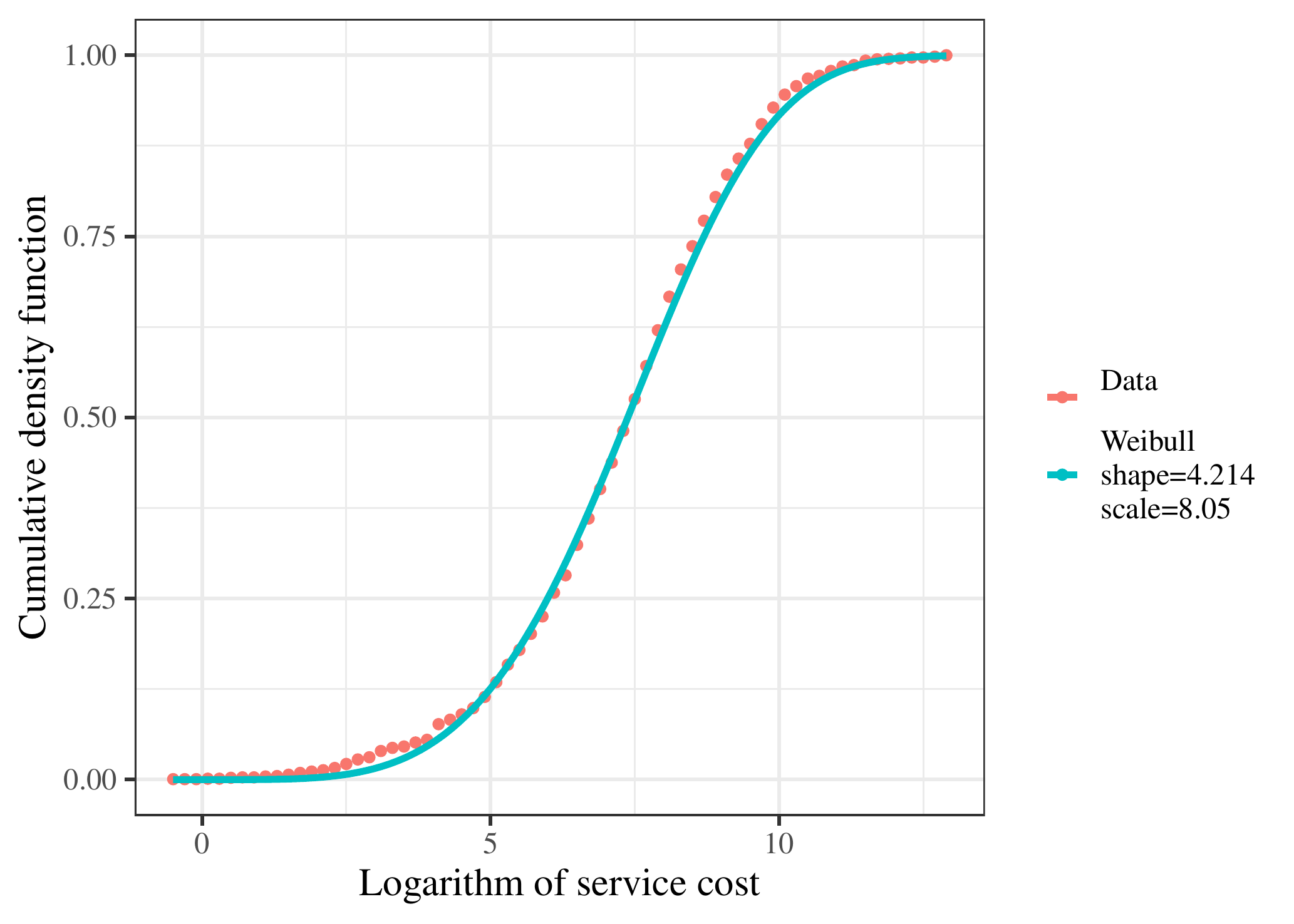}
	\caption{Fitted Distribution of empirical cost data}
	\label{fig:cdf_weibull_fit}
\end{figure}
With the parameter $\lambda$ we can simulate the aging of the system with increasing $\lambda$ over the life cycle.  The following graph simulates the risk premium in dependence of the increasing $\lambda$. 
As we have an expectation value of the life cycle cost of 8195.47 EUR the risk premium of 3000  is a significant additional amount. The reason is of course the high variance of the life cycle cost. 
\begin{figure}[H]
	\includegraphics[width=0.80\textwidth]{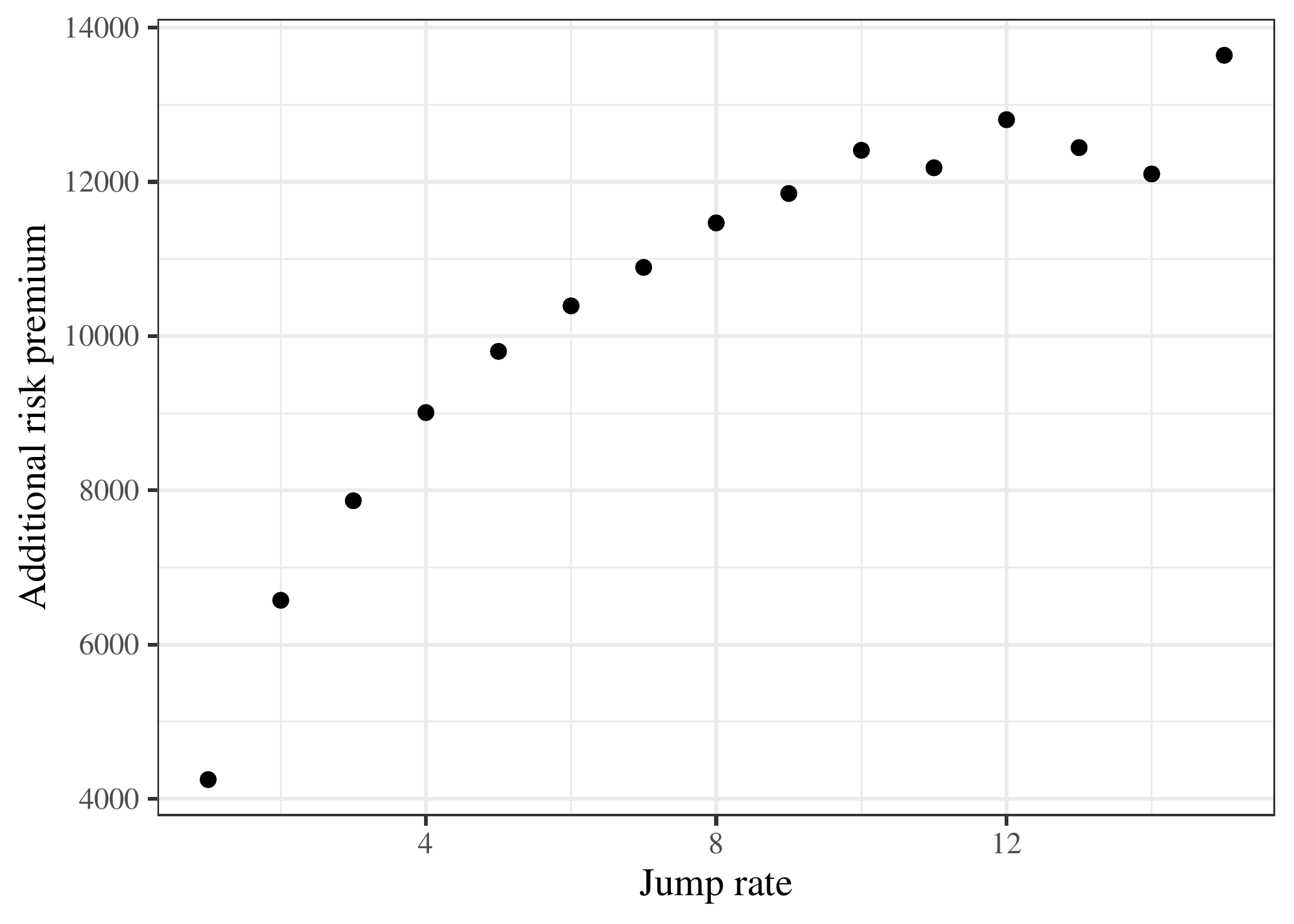}
	\caption{Risk Premium in dependence of the Jump Rate $\lambda$}
	\label{fig:Simulationen/optionPrice_overLambda}
\end{figure}
For the total price of the PBC we have to add the risk premium to the expectation value. This is shown in Fig. 11. 
\begin{figure}[H]
	\includegraphics[width=0.90\textwidth]{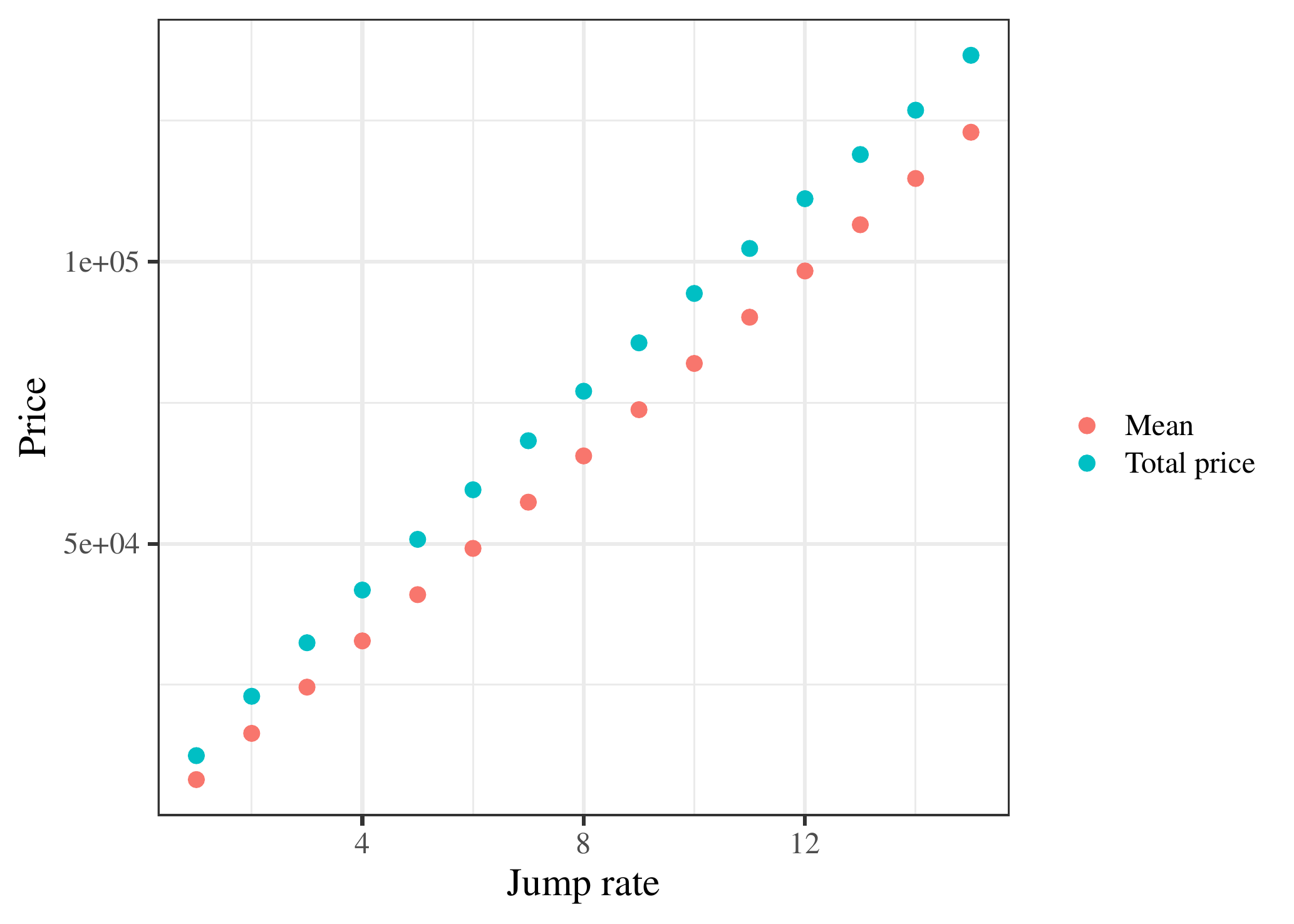}
	\caption{Mean and Total Price for the Performance Contract of a Drive System}
	\label{fig:premium_overLambda}
\end{figure}
In contrast to the Gaussian distributed cost the risk behavior for small $\lambda$ is totally different. With increasing $\lambda$ the premium cost is increasing because of the extreme fat tail of the empirical data. For the risk management it is therefore very important to know the cost distribution of the service cost in order to hedge and price the risk. Our method can be used in order to estimate the financial risk that entering the new business model. In general the analysis emphasize again that the knowledge of the jump cost distribution is the most important information for a comprehensive risk management. Nowadays this kind of  service data can be collected online with condition monitoring systems which upload the data into special cloud services. Then an automated stochastic analysis of the history could be performed. This enables a transparent and always updated financial risk management for such complex systems.
\section {Conclusion and Outlook}
The paper presents for the first time a general approach to adapt methods of asset pricing for new business models as Performance Contracts. The risk management for such new instruments are the key for best practice management. The established methods for asset pricing are now used in a different application because we apply it to operational risks and their financial impact. The main difference to option pricing is that the underlying cost processes are based on directly measurable cost functions as the life cycle cost. Due to the digitalization of condition monitoring the empirical data of such cost are online available with increasing data quality. We show how the operational risks are integrated into a financial models. Management decisions to enter new business models should be  based on simulation with real cost data of installed products. Our results shows a significant difference for the risk between the simple Gaussian distribution and real data. Simulation allows now to quantify these risk.   Based on this fundamental approach the development of new financial instrument for hedging such contracts is possible. These models open also new areas of financial activities and development in order to support further digitalization and Industry 4.0. An natural extension of our model is the inclusion of the payment conditions dependent on key performance indicators. In a certain sense this is very analog to the calculation for exotic options as barrier options. In this case the payoff function will be path dependent  as a model for the specific payment conditions of a PBC. In summary new business models will play a key role for the transformation of future business.            
\newpage
\printbibliography
\end {document}